# Topological Surface States Originated Spin-Orbit Torques in Bi$_2$Se$_3$

Yi Wang,[1] Praveen Deorani,[1] Karan Banerjee,[1] Nikesh Koirala,[2] Matthew Brahlek,[2] Seongshik Oh,[2] and Hyunsoo Yang[1,*]

[1]*Department of Electrical and Computer Engineering, National University of Singapore, 117576, Singapore*
[2]*Department of Physics & Astronomy, Rutgers Center for Emergent Materials, Institute for Advanced Materials, Devices and Nanotechnology, The State University of New Jersey, New Jersey 08854, USA*

Three dimensional topological insulator bismuth selenide (Bi$_2$Se$_3$) is expected to possess strong spin-orbit coupling and spin-textured topological surface states, and thus exhibit a high charge to spin current conversion efficiency. We evaluate spin-orbit torques in Bi$_2$Se$_3$/Co$_{40}$Fe$_{40}$B$_{20}$ devices at different temperatures by spin torque ferromagnetic resonance measurements. As temperature decreases, the spin-orbit torque ratio increases from ~ 0.047 at 300 K to ~ 0.42 below 50 K. Moreover, we observe a significant out-of-plane torque at low temperatures. Detailed analysis indicates that the origin of the observed spin-orbit torques is topological surface states in Bi$_2$Se$_3$. Our results suggest that topological insulators with strong spin-orbit coupling could be promising candidates as highly efficient spin current sources for exploring next generation of spintronic applications.

*eleyang@nus.edu.sg

The realization of functional devices such as the non-volatile memories and spin logic applications is of key importance in spintronic research [1]. The functions of these magnetic devices require highly efficient magnetization manipulation in a ferromagnet (FM), which can be achieved by an external magnetic field or a spin polarized current by spin transfer torque (STT). Recent advances have demonstrated that pure spin currents resulting from charge currents via spin-orbit coupling in heavy metals, such as Pt [2-7], Ta [8-10], and W [11], can produce strong spin-orbit torques on the adjacent magnetic layers. The reported amplitude of spin Hall angles (i.e. efficiency of spin-orbit torques) in Pt and Ta is in the range of ~ 0.012 to ~ 0.15, and in W is ~ 0.33. The exploration for new materials exhibiting new physics and possessing an even higher conversion efficiency between the charge current density ($J_c$) and spin current density ($J_s$) is crucial to exploit next generation spintronic devices.

The three dimensional (3D) topological insulators (TI) are a new class of quantum state of materials that have an insulating bulk and spin-momentum-locked metallic surface states [12-14]. They exhibit strong spin-orbit coupling and are expected to show a high charge to spin current conversion efficiency. So far, by extensively employing angle-resolved photoemission spectroscopy (ARPES) and spin-resolved ARPES, the Dirac cones and the helical spin polarized topological surface states (TSS) have been observed and the topological nature has been confirmed in TIs [15,16]. The surface state dominant conduction has also been confirmed by thickness dependent transport measurements in $Bi_2Se_3$ [17].

The TSS in TI is immune to the nonmagnetic impurities due to the time reversal symmetry protection. Although a gap opening in the TSS dispersion was reported in $Bi_2Se_3$ doped with Fe in the bulk [18], most recently reports have confirmed that the TSS is intact in Bi$_2$Se$_3$ covered with Fe [19,20] or Co [21] with in-plane magnetic anisotropy. The spin dependent transport is known to be significant near the Fermi level in the Bi$_2$Se$_3$ surface states. However, limited spin dependent transport experiments have been focused on TI/FM heterostructures. Only recently, spin-orbit effects have been reported by spin pumping measurements [22-24] and magnetoresistance measurements [25,26]. Direct charge current induced spin-orbit torque on the FM layer has been demonstrated by spin torque ferromagnetic resonance (ST-FMR) measurement only at room temperature [27] and magnetization switching at cryogenic temperature [28]. It is known that for Bi$_2$Se$_3$ the bulk channel provides an inevitable contribution to transport at room temperature and may diminish the signals of spin-orbit torques arising from surface states. At low temperatures, however, the surface contribution should become significant [17], and spin-orbit torques in TI/FM heterostructures should be enhanced [28].

In this work, we adopt extensively studied Bi$_2$Se$_3$ as the TI layer and investigate the temperature dependence of charge-spin conversion efficiency, spin-orbit torque ratio ($\theta_\parallel = J_s/J_c$), by the ST-FMR technique in Bi$_2$Se$_3$/Co$_{40}$Fe$_{40}$B$_{20}$ heterostructures. In this structure, the spin currents generated from charge currents flowing in Bi$_2$Se$_3$ are injected into ferromagnetic Co$_{40}$Fe$_{40}$B$_{20}$ layer and exert torques on it. It must be pointed out that the spin-orbit torques could be attributed to either the spin Hall effect (SHE) in the Bi$_2$Se$_3$ bulk, Rashba-split states at the interface [29-31], or Bi$_2$Se$_3$ topological surface states [23,27,28,31]. We find that $\theta_\parallel$ drastically increases when the temperature decreases to ~ 50 K. As the temperature decreases furthermore, $\theta_\parallel$ reaches up to ~ 0.42, which is ~ 10 times larger than that at 300 K. In addition, a significant out-of-plane torque is extracted at low temperatures. We argue that our observations could be correlated with the TSS in our $Bi_2Se_3/Co_{40}Fe_{40}B_{20}$ heterostructures.

20 quintuple layer (QL, 1 QL ≈ 1 nm) of $Bi_2Se_3$ films are grown on $Al_2O_3$ (0001) substrates using a custom designed SVTA MOSV-2 molecular beam epitaxy (MBE) system with a base pressure < 3 × $10^{-10}$ Torr. The detailed procedures for $Bi_2Se_3$ thin film growth can be found in previous reports [17,32]. The temperature dependent resistivity of $Bi_2Se_3$ film is measured by four probe method. Figure 1(a) shows a typical characteristic of $Bi_2Se_3$ that the sheet resistivity decreases as temperature decreases and then saturates at temperature < 30 K [17,33]. High resistivity $Co_{40}Fe_{40}B_{20}$ (CFB) is chosen as the FM layer in order to minimize the current shutting effect thru the FM layer. We have prepared five $Bi_2Se_3$/CFB ($t$) samples (thickness $t$ = 1.5, 2, 3, 4 and 5 nm) and measured their magnetization response as a function of external magnetic field as plotted in Fig. 1(b). From the inset of Fig. 1(b), the CFB dead layer in $Bi_2Se_3$/CFB samples is estimated to be 1.36 nm, similar to a recent report in which the Co dead layer at the interface of $Bi_2Se_3$/Co is ~ 1.2 nm [34].

The ST-FMR devices are fabricated by the following process. First, a 5 nm CFB layer is sputtered onto the $Bi_2Se_3$ film at room temperature with a base pressure of 3×$10^{-9}$ Torr followed by a MgO (1 nm)/$SiO_2$ (3 nm) capping layer to prevent CFB from oxidation. Then the film is patterned into rectangular shaped microstrips (dotted blue line) with dimensions of $L$ (130 μm) × $W$ (10 – 20 μm) by photolithography and Ar ion milling as shown in Fig. 2(a). In the next step, coplanar waveguides (CPWs) are fabricated. Different gaps (10 – 55 μm) between ground (G) and signal (S) electrodes are designed to tune the device impedance ~ 50

Ω. A radio frequency (RF) current ($I_{rf}$) with frequencies from 7 to 10 GHz and a nominal power of 15 dBm from a signal generator (SG, Agilent E8257D) is applied to the $Bi_2Se_3$/CFB bilayer via a bias-tee, and the ST-FMR signal ($V_{mix}$) is detected simultaneously by a lock-in amplifier. An in-plane external magnetic field ($H_{ext}$) is applied at a fixed angle ($\theta_H$) of 35º with respect to the microstrip length direction [6]. We present the data from three different devices, denoted as D1, D2 and D3.

Figure 2(b) shows the measured ST-FMR signals from D1 at different temperatures ranging from 20 to 300 K. $V_{mix}$ can be fitted by a sum of symmetric and antisymmetric Lorentzian functions, $V_{mix} = V_s F_{sym}(H_{ext}) + V_a F_{asym}(H_{ext})$ [3,6,27]. From fitting, the symmetric component $V_s$ (corresponding to in-plane torque $\tau_\parallel$ on CFB) and antisymmetric component $V_a$ (corresponding to total out-of-plane torque $\tau_\perp$) can be determined, simultaneously.

The spin-orbit torque ratio from ST-FMR measurements can be characterized by two methods. One is to obtain $\theta_\parallel$ from the analysis of $V_s/V_a$ via $\theta_\parallel = (V_s/V_a)(e\mu_0 M_s t d/\hbar)[1+(4\pi M_{eff}/H_{ext})]^{1/2}$ [3], where $t$ and $d$ represent the thickness of the CFB and $Bi_2Se_3$ layer, respectively. $M_s$ is the saturation magnetization of CFB and $M_{eff}$ is the effective magnetization. This method (denoted as 'by $V_s/V_a$' hereafter) is to date widely used in ST-FMR measurements of heavy metals Pt (or Ta)/FM bilayers [3,6,8]. However, one assumption of this method is that the $V_a$ is only attributed to the Oersted field induced out-of-plane torque. However, in the case of a TI, the TSS in TI and/or Rashba-split states at the interface could also contribute to $V_a$, therefore, we cannot estimate the actual $\theta_\parallel$ value by $V_s/V_a$. On the other hand, the second method can avoid such an issue by analyzing only the symmetric component $V_s$ (denoted as 'by $V_s$ only' hereafter) using the following equations:

$$V_s = -\frac{I_{rf}\gamma \cos\theta_H}{4}\frac{dR}{d\theta_H}\tau_{\parallel}\frac{1}{\Delta}F_{sym}(H_{ext}), \quad \sigma_s = J_s/E = \tau_{\parallel}M_s t/E, \quad \text{and} \quad \theta_{\parallel} = \sigma_s/\sigma \quad [6,27],$$ where $I_{rf}$ is the RF current flowing through the device, $dR/d\theta_H$ is the angular dependent magnetoresistance at $\theta_H = 35°$, $\Delta$ is the linewidth of ST-FMR signal, $F_{sym}(H_{ext})$ is a symmetric Lorentzian, $\tau_{\parallel}$ is the in-plane spin-orbit torque on unit CFB moment at $\theta_H = 0°$, $\sigma_s$ is the $Bi_2Se_3$ spin Hall conductivity, $\sigma$ is the $Bi_2Se_3$ conductivity, and $E$ is the microwave field across the device. The second method avoids the possible contamination to $\theta_{\parallel}$ arising from $V_a$, therefore we can extract the $\theta_{\parallel}$ values in $Bi_2Se_3$ by analyzing only $V_s$. At the same time, the total out-of-plane torque $\tau_{\perp}$ can be derived by using

$$V_a = -\frac{I_{rf}\gamma \cos\theta_H}{4}\frac{dR}{d\theta_H}\tau_{\perp}\frac{[1+(\mu_0 M_{eff}/H_{ext})]^{1/2}}{\Delta}F_{asym}(H_{ext}) \quad [27],$$ where $F_{asym}(H_{ext})$ is an antisymmetric Lorentzian.

Figure 3(a-b) show the $\tau_{\parallel}$ and $\tau_{\perp}$ as functions of temperature, respectively, using the 2$^{nd}$ method. Here, the $\tau_{\parallel}$ ($\tau_{\perp}$) represents the mean value for different RF frequencies. At 300 K, the $\tau_{\parallel}$ is ~ 0.43 Oe for D1 (~ 0.84 Oe for D2 and ~ 0.48 Oe for D3). As the temperature decreases from 300 to 100 K, $\tau_{\parallel}$ for all three devices gradually increases. At ~ 50 K, $\tau_{\parallel}$ shows a steep increase and finally reaches ~ 5.25 Oe for D1 (~ 4.11 Oe for D2 and ~ 2.26 Oe for D3), which is ~ 10 times larger than that at 300 K. It is noteworthy that the observed drastic temperature dependent behavior of $\tau_{\parallel}$ is different from the recently reported results in heavy metals such as Ta [10,35] as well as Pt [6,36,37], where the damping-like torque (equivalent to $\tau_{\parallel}$ here), often argued to arise mainly from the SHE, shows a weak temperature dependence. This difference indicates the SHE mechanism may not account for the observed $\tau_{\parallel}$ in our $Bi_2Se_3$/CFB. Moreover, the $\tau_{\perp}$ shows a similar temperature dependent behavior as $\tau_{\parallel}$

shown in Fig. 3(b). It is worth noting that the difference in $\tau_{\|}$ (and $\tau_{\perp}$) among D1, D2 and D3 can be attributed to the slight variation of the $Bi_2Se_3$/CFB interface during the fabrication process considering recent challenges in TI film growth and device fabrication. However, a qualitatively similar temperature dependence of torques is observed in all devices.

The $\theta_{\|}$ values as a function of temperature determined by above two methods have been shown in Fig. 3(c). From analysis by $V_s$ only, $\theta_{\|}$ is ~ 0.047 for D1 (~ 0.113 for D2 and ~ 0.072 for D3) at 300 K, and increases to ~ 0.158 for D1 (~ 0.225 for D2 and ~ 0.149 for D3) as temperature decreases to 100 K. In this temperature range (100 - 300 K), $\theta_{\|}$ has similar amplitudes as the spin Hall angle in heavy metals such as Pt, Ta, and W [3,8,11,42-44]. However, $\theta_{\|}$ increases sharply as temperature decreases to ~ 50 K and reaches maximum values of ~ 0.42 for D1 (~ 0.44 for D2 and ~ 0.30 for D3) at lower temperatures, respectively. Remarkably, $\theta_{\|}$ increases ~ 10 times compared to that at 300 K for D1. Similarly, from the analysis by $V_s/V_a$, $\theta_{\|}$ also shows an abrupt increase as temperature decreases to ~ 50 K in Fig. 3(c). It is worth noting that we use the effective CFB thickness of $t$ = 3.64 nm due to the dead layer for $\theta_{\|}$ estimation by $V_s/V_a$ at different temperatures. Interestingly, as shown in Fig. 3(d), the ratio of [$\theta_{\|}$ (by $V_s$ only) – $\theta_{\|}$ (by $V_s/V_a$)]/$\theta_{\|}$ (by $V_s/V_a$) obtained by two different methods increases as temperature decreases and becomes more significant below ~ 50 K, as discussed later.

In the context of spin Hall mechanism, the spin Hall angle ($\theta_{sh}$) is found to be almost independent of temperature from Pt [6,36], Ta [45], $Cu_{99.5}Bi_{0.5}$, and $Ag_{99}Bi_1$ [46], which is attributed to the extrinsic mechanisms. In some cases, $\theta_{sh}$ shows a gradual increase as the temperature decreases, which behaves as a typical intrinsic mechanism based on the degeneracy of d-orbits by spin-orbit coupling [47,48]. In contrast, in our $Bi_2Se_3$/CFB, the spin-orbit torque ratio ($\theta_\parallel$) shows an abrupt and nonlinear increase as temperature decreases, especially below ~ 50 K. Therefore, the SHE from the $Bi_2Se_3$ bulk is probably not the dominant mechanism for our observation of temperature dependent spin-orbit torque (ratio) in $Bi_2Se_3$/CFB. From the measured ST-FMR signals as shown in Fig. 2(b), we also find that the Rashba-split state at the $Bi_2Se_3$/CFB interface is not the main mechanism for our observations, since the Rashba-split states lead to opposite direction (and sign) of charge current-induced spin polarization (and $\theta_\parallel$) on the basis of the spin structure [27,31]. Instead, we ascertain that the direction of in-plane spin polarization to the electron momentum in our $Bi_2Se_3$/CFB is consistent with expectations of the TSS of TIs (spin-momentum locking) [12-14,27,31,37]. From further analysis [37], we have found that in our devices a large portion of the charge current flows through the TSS in $Bi_2Se_3$. The effective $\theta_\parallel$ attributed to only TSS is in the range from ~ 1.62 ± 0.18 to ~ 2.1 ± 0.39.

As mentioned before, the temperature dependent $\theta_\parallel$ obtained from the above two methods shown in Fig. 3(c) should not show any difference, if $V_a$ is attributed to only the charge current induced Oersted field. Therefore, the observed difference implies the existence of other contributions to $V_a$ (i.e. to $\tau_\perp$). For the $Bi_2Se_3$/CFB system, the difference can be attributed to the TSS in $Bi_2Se_3$ [23,27,28,31] and/or Rashba-split states at the $Bi_2Se_3$/CFB interface [29-31]. We analyze $\Delta\tau = \tau_\perp - \tau_{Oe}$ as the other contributions to the out-of-plane torque, where $\tau_\perp$ is the total out-of-plane torque as shown in Fig. 3(b), and $\tau_{Oe}$ is a partial out-of-plane torque from charge current (flowing in $Bi_2Se_3$) induced Oersted field. By using the measured $\theta_\parallel$ by $V_s$ only, we can deduce $\tau_{Oe}$ and thus $\Delta\tau$ by

$$\theta_{\parallel} = (V_s/V_a^*)(e\mu_0 M_s td/\hbar)[1+(4\pi M_{eff}/H_{ext})]^{1/2}$$ , and

$$V_a^* = -\frac{I_{rf}\gamma\cos\theta_H}{4}\frac{dR}{d\theta_H}\tau_{Oe}\frac{[1+(\mu_0 M_{eff}/H_{ext})]^{1/2}}{\Delta}F_{asym}(H_{ext})$$ [3,27], where $V_a^*$ is the

equivalent antisymmetric component only due to the current induced Oersted field ($\tau_{Oe}$). As shown in Fig. 4(a), the out-of-plane torque ($\Delta\tau$) in all three devices becomes much larger at low temperatures < 50 K, compared to the $\Delta\tau$ at high temperatures (100 – 300 K). Consequently, we can obtain the out-of-plane spin-orbit torque ratio ($\theta_\perp$) as a function of temperature by using the same method by which we deduce $\theta_\parallel$ from $\tau_\parallel$ above. As shown in Fig. 4(b), we find that $\theta_\perp$ in all three devices also becomes more significant at low temperatures (< 50 K). More interestingly, the $\theta_\perp$ almost has the same order of magnitude compared to $\theta_\parallel$.

We now discuss the origin of the out-of-plane torque. As has been reported recently, a Rashba-split surface state in two dimensional electron gas (2DEG) coexists with TSS in the Bi$_2$Se$_3$ surface due to the band bending and structural inversion asymmetry [29,30,49-52]. The Rashba effective magnetic field can be written as $H_T = \alpha_R/\hbar(\hat{z}\times k)$ [49-51], where $\hat{z}$ is a unit vector normal to film plane, $k$ is the average electron Fermi wavevector, and $\alpha_R$ is a characteristic parameter of the strength of Rashba splitting in 2DEG. Since the electron Fermi wavevector can be assumed to show a weak temperature dependence and the $\alpha_R$ decrease as temperature decreases in a typical 2DEG [53,54], $H_T$ is expected to decrease as temperature decreases in these semiconductor systems. In addition, the similar temperature dependent behavior of $H_T$ has been recently reported in Ta/CoFeB heterostructures, where $H_T$ decreases and eventually almost reaches to zero at low temperatures [10,35]. However, the observed $\Delta\tau$ (equivalent to $H_T$) in our Bi$_2$Se$_3$/CFB presents the opposite temperature dependent behavior which is not in line with the reports about Rashba induced torques. Therefore, we conclude that the Rashba-split surface state in 2DEG of $Bi_2Se_3$ is not the main mechanism for the out-of-plane torque ($\Delta\tau$).

On the other hand, a possible out-of-plane spin polarization in the TSS has been theoretically predicted [55,56] and experimentally observed in $Bi_2Se_3$ [57,58], which is attributed to the hexagonal warping effect in the Fermi surface [55,59]. This out-of-plane spin polarization in the TSS can account for the observed $\Delta\tau$ especially in the low temperature range (< 50 K) and the $\Delta\tau$ adds to the $\tau_{Oe}$ [27,31]. Moreover, as shown in Fig. 3(a) and 4(a), the out-of-plane torque ($\Delta\tau$) has the same order of magnitude comparable to in-plane torque ($\tau_\parallel$) below 50 K ($\Delta\tau/\tau_\parallel \sim 60\%$) [37], which is in agreement with the behavior of hexagonal TSS in TI [55,56]. With the analysis from different aspects, our findings especially in the low temperature range (< 50 K) indicate a TSS origin of spin-orbit torques in $Bi_2Se_3$/CFB.

In summary, we have studied the temperature dependence of spin-orbit torques in $Bi_2Se_3$/CoFeB heterostructures. As temperature decreases, the spin-orbit torque ratio increases drastically and eventually reaches a maximum value of ~ 0.42, which is almost 10 times larger than that at 300 K. A significant out-of-plane torque ($\Delta\tau$), in addition to charge current induced Oersted field torque ($\tau_{Oe}$), can be observed below 50 K. The observed spin-orbit torques are attributed to the topological surface states in $Bi_2Se_3$. Our results suggest that topological insulators with strong spin-orbit coupling and spin-momentum locking are promising spin current sources for next generation of spintronic devices.

This work was partially supported by the National Research Foundation (NRF), Prime Minister's Office, Singapore, under its Competitive Research Programme (CRP award no. NRFCRP12-2013-01), the Ministry of Education-Singapore Academic Research Fund Tier 1 (R-263-000-A75-750) & Tier 2 (R-263-000-B10-112), Office of Naval Research (N000141210456), and by the Gordon and Betty Moore Foundation's EPiQS Initiative through Grant GBMF4418.

**Figure captions**

FIG. 1. (a) Temperature dependent sheet resistivity of $Bi_2Se_3$ films (20 QL). (b) The magnetization versus field ($H$) for $Bi_2Se_3$ (20 QL)/CFB ($t$) (*nominal* thickness $t$ =1.5, 2, 3, 4 and 5 nm) at room temperature. The inset shows the magnetization per unit area versus CFB thickness.

FIG. 2. (a) The schematic diagram of the ST-FMR measurement, illustrating a bias-tee, lock-in amplifier, RF signal generator (SG), and ST-FMR device with a $Bi_2Se_3$/CFB (5 nm). Micro-strip is denoted by a dashed blue rectangle. (b) The measured ST-FMR signals from a $Bi_2Se_3$/CFB (5 nm) device (D1) at different temperatures.

FIG. 3. Temperature dependence of (a) $\tau_\parallel$, (b) $\tau_\perp$, (c) $\theta_\parallel$, and (d) [$\theta_\parallel$ (by $V_s$ only) − $\theta_\parallel$ (by $V_s/V_a$)]/[$\theta_\parallel$ (by $V_s/V_a$)] in $Bi_2Se_3$/CFB (5 nm) for D1, D2, and D3. The $\theta_\parallel$ is analyzed by two different methods, by '$V_s$ only' and by '$V_s/V_a$'.

FIG. 4. (a) Temperature dependent out-of-plane torque ($\Delta\tau = \tau_\perp − \tau_{Oe}$) and (b) out-of-plane torque ratio ($\theta_\perp$) in $Bi_2Se_3$/CFB (5 nm) devices.

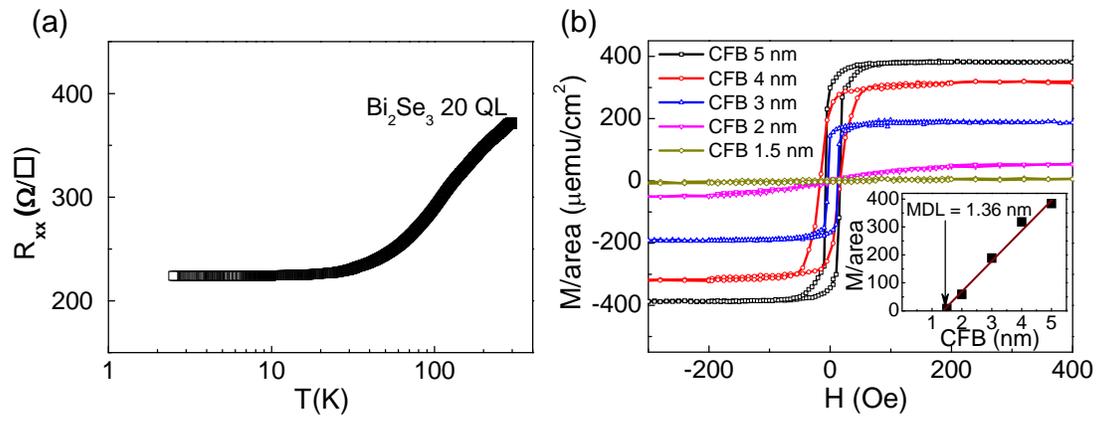

FIG. 1

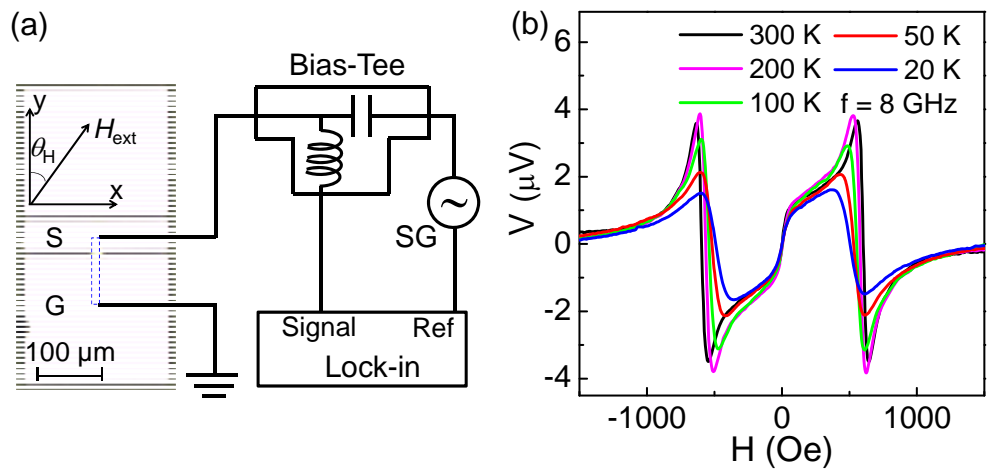

FIG. 2

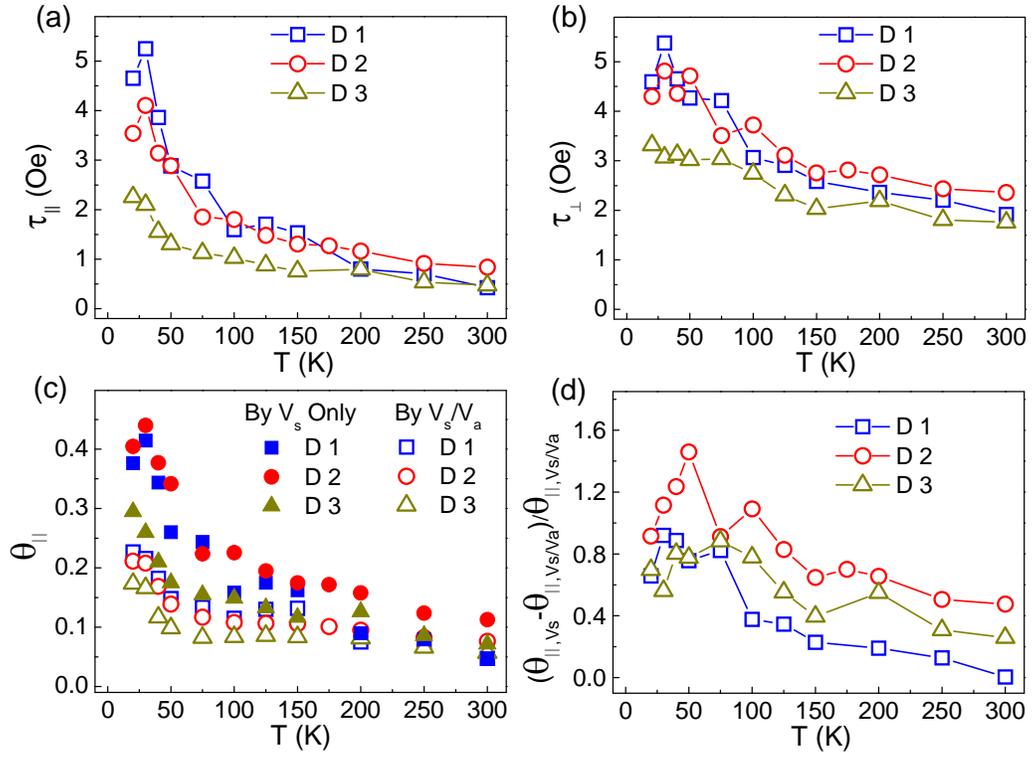

FIG. 3

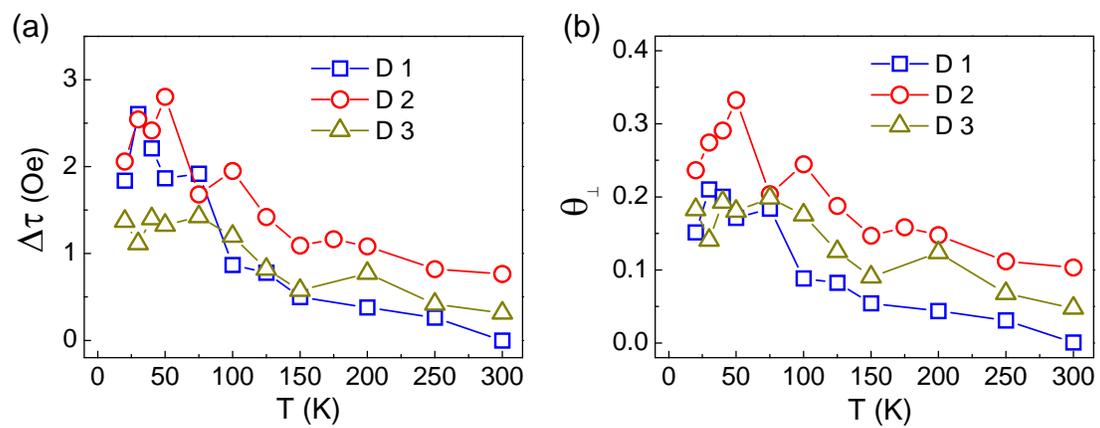

FIG. 4
18